\documentclass[sn-mathphys-num]{sn-jnl}
\usepackage{lipsum}
\usepackage{dsfont}

\usepackage{graphicx}%
\usepackage{multirow}%
\usepackage{amsmath,amssymb,amsfonts}%
\usepackage{amsthm}%
\usepackage{mathrsfs}%
\usepackage[title]{appendix}%
\usepackage{xcolor}%
\usepackage{textcomp}%
\usepackage{manyfoot}%
\usepackage{booktabs}%
\usepackage{algorithm}%
\usepackage{algorithmicx}%
\usepackage{algpseudocode}%
\usepackage{listings}%
\usepackage{filecontents}
\usepackage{caption,subcaption}
\usepackage{xfrac}
\usepackage{nicefrac}
\usepackage{color}
\usepackage{colortbl}
\usepackage{soul}
\usepackage{pgfplots}
\usepackage{ulem}
\usepackage{censor}
\usepackage{natbib}
\usepackage{cancel}

\newcommand{\RRR}{{\mathbb{R} }}

\newcommand{\Qc}{{\mathcal{Q} }}

\newcommand{\bs}[1]{\boldsymbol{#1}}

\newcommand{\qq}{\bs{q}}

\newcommand{\vv}{\bs{v}}

\newcommand{\xx}{\bs{x}}

\newcommand{\DD}{\bs{D}}

\newcommand{\JJ}{\bs{J}}
\newcommand{\KK}{\bs{K}}

\newcommand{\ttau}{\bs{\tau}}

\newcommand{\oomega}{\bs{\omega}}

\makeatletter
\renewcommand*\env@matrix[1][\arraystretch]{%
	\edef\arraystretch{#1}%
	\hskip -\arraycolsep
	\let\@ifnextchar\new@ifnextchar
	\array{*\c@MaxMatrixCols c}}
\makeatother

\definecolor{TUMBlue}{HTML}{0065BD}

\DeclareUnicodeCharacter{2212}{-}

\newcommand*{\tran}{^\mathsf{T}}

 \DeclareRobustCommand{\rchi}{{\mathpalette\irchi\relax}}
\newcommand{\irchi}[2]{\raisebox{\depth}{$#1\chi$}}

\begin{document}

\title[Identifying Visible Tissue in Intraoperative Ultrasound: A Method and Application]{Identifying Visible Tissue in Intraoperative Ultrasound: A Method and Application}

\author[1]{\fnm{Alistair} \sur{Weld}}\email{a.weld20@imperial.ac.uk}
\author[2]{\fnm{Luke} \sur{Dixon}} 
\author[3]{\fnm{Michael} \sur{Dyck}} 
\author[4]{\fnm{Giulio} \sur{Anichini}} 
\author[1]{\fnm{Alex} \sur{Ranne}}
\author[4]{\fnm{Sophie} \sur{Camp}}
\author[1]{\fnm{Stamatia} \sur{Giannarou}}

\affil[1]{\orgdiv{The Hamlyn Centre}, \orgname{Imperial College London}, \orgaddress{\country{UK}}}

\affil[2]{\orgdiv{Department Of Imaging}, \orgname{Charing Cross Hospital}, \orgaddress{\country{UK}}}

\affil[3]{\orgdiv{School Of Computation, Information and Technology}, \orgname{Technical University of Munich}, \orgaddress{\country{DE}}}

\affil[4]{\orgdiv{Department Of Neurosurgery}, \orgname{Charing Cross Hospital}, \orgaddress{\country{UK}}}


\abstract{\textbf{Purpose:} Intraoperative ultrasound scanning is a demanding visuotactile task. It requires operators to simultaneously localise the ultrasound perspective and manually perform slight adjustments to the pose of the probe, making sure not to apply excessive force or breaking contact with the tissue, whilst also characterising the visible tissue. \textbf{Method:} To analyse the probe-tissue contact, an iterative filtering and topological method is proposed to identify the underlying visible tissue, which can be used to detect acoustic shadow and construct confidence maps of perceptual salience.  \textbf{Results:} For evaluation, datasets containing both in vivo and medical phantom data are created. A suite of evaluations is performed, including an evaluation of acoustic shadow classification. Compared to an ablation, deep learning, and statistical method, the proposed approach achieves superior classification on in vivo data, achieving an $F_{\beta}$ score of 0.864, in comparison to 0.838, 0.808, 0.808. A novel framework for evaluating the confidence estimation of probe tissue contact is created. The phantom data is captured specifically for this, and comparison is made against two established methods. The proposed method produced the superior response, achieving an average normalised root mean square error of 0.168, in comparison to 1.836 and 4.542. Evaluation is also extended to determine the algorithm's robustness to parameter perturbation, speckle noise, data distribution shift, and capability for guiding a robotic scan. \textbf{Conclusion:} The results of this comprehensive set of experiments justify the potential clinical value of the proposed algorithm, which can be used to support clinical training and robotic ultrasound automation.}


\maketitle

\section{Introduction}\label{sec_intro}

Intraoperative imaging is an important tool for surgeons in decision-making during surgery. The low cost of ultrasound (US) machines and the ease of integration into surgical settings has led to enthusiasm for intraoperative ultrasound (IOUS) \cite{Sastry2017ApplicationsOU}. However, the widespread adoption of IOUS has been limited and there remains great performance variability between operators \cite{Dixon2022IntraoperativeUI}. Becoming proficient in IOUS requires extensive training and hands-on experience, due to complex visuotactile interaction, imaging artefacts, noise, machine settings, and a lack of clear global anatomical key structures, which can cause interpretation ambiguity. 

One particularly challenging aspect is balancing optimal probe-tissue contact and orientation, without causing inadvertent pressure-related damage. Contact related acoustic shadow (AS) can occur due to interposed gas or contact breaking, which reduces image reliability through tissue obscuration. This is particularly challenging in the case of brain surgery \cite{Dixon2022IntraoperativeUI}.

\begin{figure}[!htbp]
\centering
    \includegraphics[height=3.8cm,width=0.45\columnwidth]{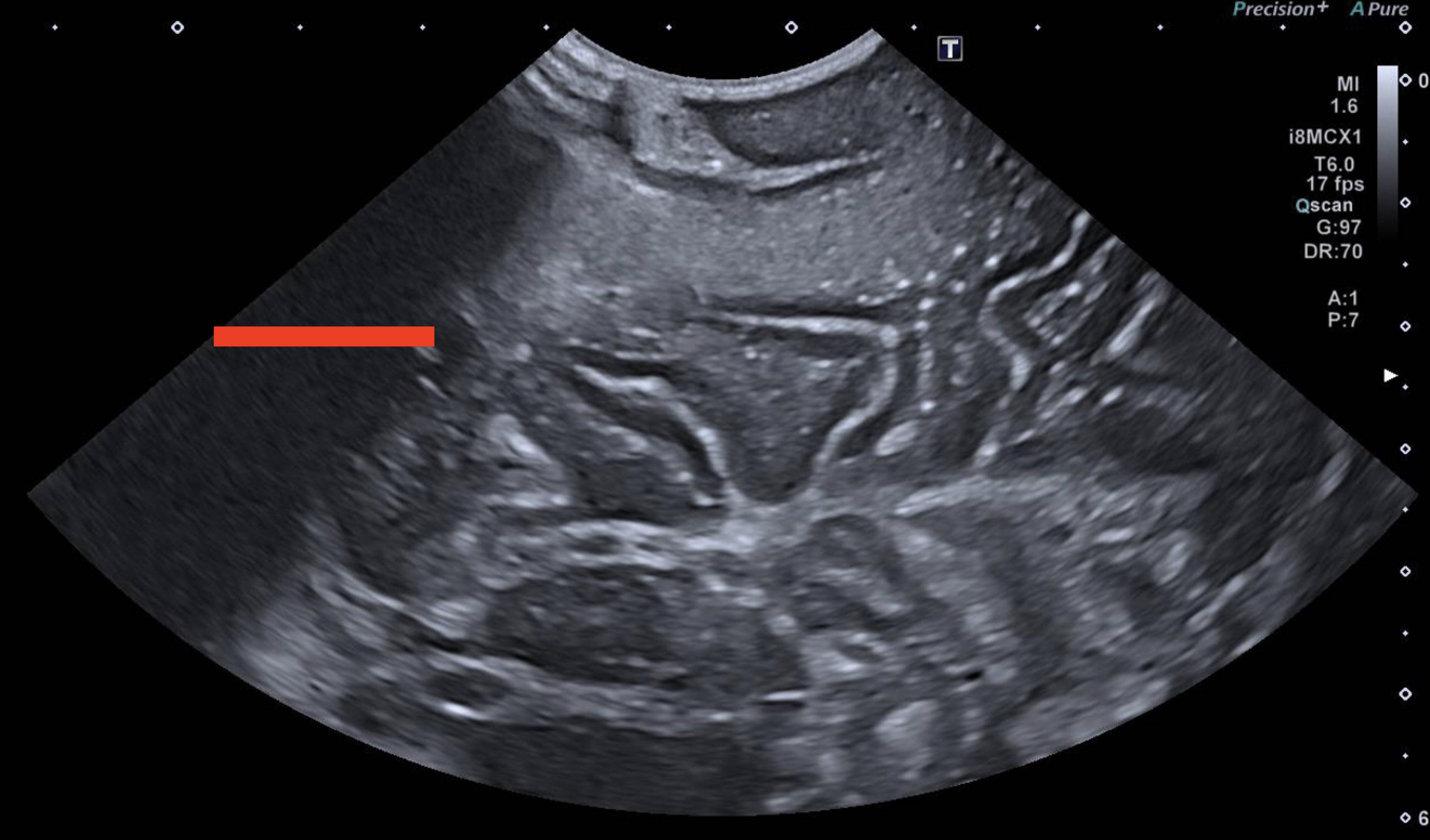}\hspace{0.em}
    \includegraphics[height=3.8cm,width=0.45\columnwidth]{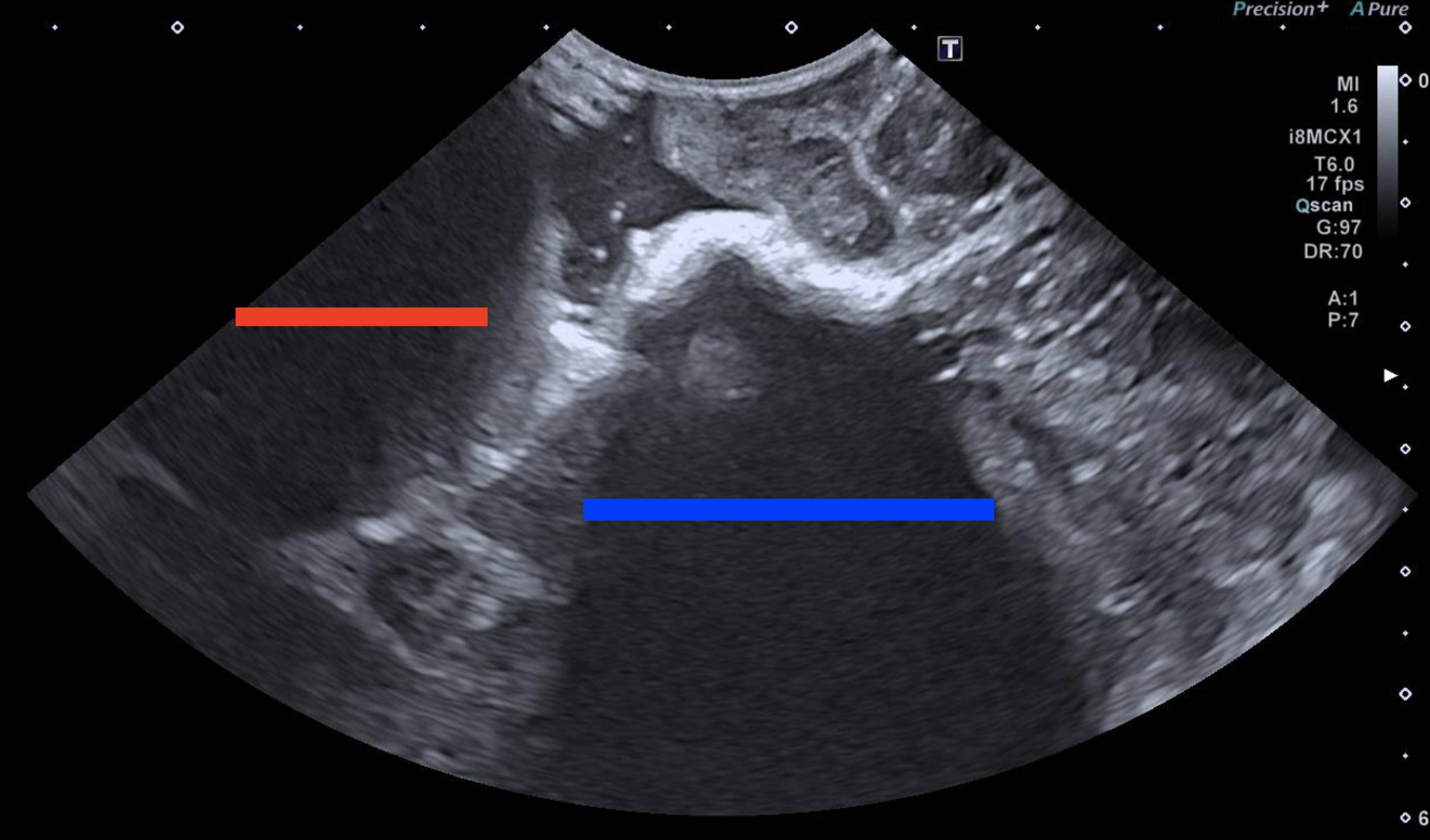}\hspace{0.5em}
   
\caption{Sample annotations of b-mode ultrasound images. The red line delineates acoustic shadowing secondary to poor contact whilst the blue line defines acoustic shadowing related to other causes (in this case echogenic blood clot in a resection cavity).}
\label{fig:blodclot}
\end{figure}

\textcolor{black}{Recent advances in IOUS have led to increased interest in algorithmic analysis of US feedback. Some methods detect shadows caused by bone \cite{Berton2016SegmentationOT, Alsinan2020BoneSS}, calcification \cite{Filho2006SegmentationOC, Lee2018AutomaticDO}, or general probe–tissue interactions \cite{Hellier2010AnAG, Meng2019WeaklySE, Melapudi2022ExploitingAP}. However, these methods do not specifically examine probe-tissue coupling, which is the focus of this work.} Fig.~\ref{fig:blodclot} presents an example of where shadowing does not always equal poor contact. To address this limitation, the algorithm in \cite{Karamalis2012USCM, Chatelain2017ConfidenceDrivenCO, Jiang2020AutomaticNP} produces a confidence map of signal attenuation, using a random walk-based method. The limiting factor for the latter method is the assumption that hyperechoic lines correlate with reduced signal attenuation. However, this is not a well-generalised assumption. In this work, we propose a method for identifying and topologically representing visible tissue in US for the detection of contact related AS and confidence in perceptual salience. Our contributions are as follows: 

\begin{enumerate}
    \item A method for identifying visible tissue within US images is proposed, using iterative filters and topological representation, that can be used for probe-tissue contact-related AS detection.

    \item A novel framework is proposed for the analysis of the confidence estimate of ultrasound probe-tissue contact.

\end{enumerate}

\section{Methodology}\label{sec_method}

\subsection{Detection of tissue features}
To detect areas of high spatial intensity variation, linear US images $I\in \mathbb{R}^{height \times width}$ are iteratively processed using Gaussian filters, with a vertical bias, following: 
\begin{equation}
\begin{gathered}
G_{f}=h_{1}*G_{f-1},\\h_{1}=(\sfrac{1}{\sqrt{2\pi\sigma_{u}}e^{-\sfrac{u^{2}}{2\sigma_{u}^{2}}}})(\sfrac{1}{\sqrt{2\pi\sigma_{v}}e^{-\sfrac{v^{2}}{2\sigma_{v}^{2}}}}), \\ 
\sigma_{u}>>\sigma_{v}, G_{1}=I
\end{gathered}
\label{eq:median}
\end{equation}
where, $G\in \mathbb{R}^{height \times width}$ denotes an output of the Gaussian filter, $f$ the iteration set, and $\sigma$ the standard deviation of the Gaussian filter stack. \textcolor{black}{The image coordinates are $u$ and $v$, where $v$ increases with signal depth along the scan lines (image columns) that are perpendicular to the transducer at the top.} The Gaussian smoothing is biased vertically, as anatomical changes are more likely to be exposed in the vertical direction. The function is linear and the iterative processing is associative $(G*h_1)*h_1 = G*(h_1*h_1)$ and therefore equal to a single convolution with a wider standard deviation $\hat{\sigma_u}\approx\sqrt{f}\sigma_u$. \textcolor{black}{Direct edge-detection methods (e.g., Laplacian or Sobel) are not robust to noise and do not yield a smooth response.} Each output $G_{f}$ of the Gaussian filter sequence is stacked and the standard deviation is taken for each pixel across the stack following: 
\begin{equation}
S_{u,v} = \hat{G}*\sqrt{\frac{\sum_{f=1}^{len(f)+1} (G_{f,u,v}-  \frac{\sum_{f=1}^{len(f)+1}G_{f,u,v}}{len(f)+1})^{2}}{len(f)+1}}
\label{eq:std}
\end{equation}
where, $S\in \mathbb{R}^{height \times width}$ is the output of the standard deviation. $\hat{G}$ is an extra smoothing filter for the standard deviation output with $\sigma=1$, to remove outliers. 

Points with high spatial variation correspond to areas that have been most affected by the smoothing process. Hence, these points are separated by thresholding the standard deviation output following:
\begin{equation}
\rchi = \{ (u,v) \mid S_{u,v} \geq \gamma \}, \gamma=round(\sigma_u \times \sqrt{len(f)})
\label{eq:thresholding}
\end{equation}
the $\gamma$ formulation is elaborated on in Sec.~\ref{sec:param_pert}. Points of high spatial variation are represented as a 2D point cloud $\rchi$. 

To reduce the computational load for building the topology and to remove outliers, the point cloud is sparsified utilising an inverse Gaussian density map $\lambda\in \mathbb{R}^{height\times width}$ following:

\begin{gather}
\textcolor{black}{\lambda = \phi_{1} \;+\; \phi_{2}\!\Bigl( 1 - \frac{\mathbf{1}_{\phi_{2}\times\phi_{2}}}{\phi_{2}^{2}} \,\ast\, \frac{\bigl(h_{2}\ast (S>\gamma)\bigr)^2} {\max \bigl(\bigl(h_{2}\ast (S>\gamma)\bigr)^2\bigr)} \Bigr),} \\
\textcolor{black}{h_{2}(u,v) = \frac{1}{2\pi\,\sigma_{w}^{2}} \exp\!\bigl(-\tfrac{u^{2}+v^{2}}{2\,\sigma_{w}^{2}}\bigr)}
\label{eq:dense}
\end{gather}

\textcolor{black}{where $\phi$ denotes the weighting parameters, \(\mathbf{1}_{\phi_{2}\times\phi_{2}}\) is a matrix \(\phi_{2}\times\phi_{2}\) of ones, \(h_{2}\) is the Gaussian kernel, and \((S>\gamma)\) is a 2D binary array.} The point cloud is then iterated over, where Eq.~\ref{eq:sparse} is repeated until all points are separated by their corresponding $\lambda_{a}$ distance threshold:

\begin{gather}
    A := \{ a \in \rchi \mid \exists b \in \rchi, b \neq a : dist(a, b) \leq \lambda_a \}, \nonumber \\
 \rchi - A := \{ a \in \rchi \mid \forall b \in \rchi, b \neq a : dist(a, b) > \lambda_a \}   
 \label{eq:sparse}
\end{gather}
where, $a,b$ indexes elements in the point cloud set, and $dist()$ gives the Euclidean distance between two points (pixels).

To generate the topological representation of the visible tissue, a Vietoris–Rips complex is then constructed on the sparse point cloud, defined: 
\begin{equation}
\Psi := \{ d \subseteq \rchi \mid \forall a,b \in d : dist(a,b) \leq \epsilon \}
\label{eq:vr}
\end{equation}
where $\epsilon$ defines the maximum distance threshold. The Vietoris–Rips complex method was chosen over alternatives such as the Delaunay method because of its simplicity and robustness to outliers.

\subsection{Artefact Detection and Confidence Map Generation}\label{subsec:artefact_detection}

\textcolor{black}{A 2D simplex \(\delta\) is made up of three points within the complex \(\Psi\), whose pairwise distances all lie within the threshold \(\epsilon\); let \(\Delta \subseteq \Psi\) be the set of all 2D simplices. For each scan line \(\xi_{sl}\) (where \(sl \in [1 \dots \text{width}]\)), we assign a value of 1 (AS) or 0 (visible tissue). Contact-related AS is defined for a scan line if the number of overlapping 2D simplices in \(\Delta\) along that line is below a specified threshold \(\tau\).}

The confidence map is a pixel-wise quantification of the perceptual salience, which represents the resolution and contrast of the detected visible tissue. A high score indicates high-quality tissue contrast and resolution. For every pixel in the image, the number of 2D simplexes that overlap is counted, creating an occurrence map $O$. The confidence map is constructed by taking the log of the occurrence map, $\hat{M}=log(O+1)$.

\section{Resources And Evaluation Setup}\label{sec_exp}

\subsection{Intraoperative Data}\label{data}

\subsubsection{Data curation} For the AS classification, a data set containing 51 images from 11 different patients was collected. The images were sourced from routine pseudo-anonymised 2D US scans, acquired in vivo during brain surgery, using a Canon i900 US machine, using both a linear and a convex probe. The data was retrospectively reviewed and annotated by a consultant neuroradiologist experienced in neurooncological iUS. 
The images exhibit varying degrees of AS artefacts secondary to poor probe-tissue contact. The study had full local ethical approval from the HRA and Health and Care Research Wales (HCRW) authorities\footnote[1]{Study title - US-CNS: Multiparametric Advanced Ultrasound Imaging of the Central Nervous System Intraoperatively and Through Gaps in the Bone, IRAS project ID: 275556, Protocol number: 22CX7609, REC reference: 22/WA/0259, Sponsor: Research Governance and Integrity Team (RGIT).}. 
For standardisation, all convex images were linearised using a standard approach of taking the inner trapezoid and interpolating. All linear images have the shape $300\times600$. To reduce the influence of near-field artefacts, the top 100 rows are cropped from all linearised US images. This choice was based on observations from the in vivo dataset and the maximum point where the near-field artefacts, caused by factors that included sheathing, gel, and cranial cavity interaction, ended. 

For performance evaluation, three-fold cross-validation was used, the data set is split into a seen and unseen subset, with the division done on the patient level.

\subsubsection{Speckle noise robustness evaluation setup}

Speckle noise is the most prominent artefact, created by the interaction of signals with biological tissue, which can cause distortion of anatomical details and edges. As proposed in \cite{Jain2018FundamentalsOD}, the observed US image $I$ can be defined as the product of the underlying clean image $\hat{I}$ affected by multiplicative $\xi_m$ and additive $\xi_a$ noise, given by:
\begin{equation} 
I = \hat{I} \odot \xi_m \oplus \xi_a
\label{eq:speck}
\end{equation}
where, it is commonly assumed that $||\xi_a||^2\ll||\hat{I} \odot \xi_m||^2$ \cite{Zong1998SpeckleRA}. Commonly, the additive term is ignored due to the lesser impact. Assuming sampling from a generalised Gamma distribution (GGD) \cite{Michailovich2006DespecklingOM} with the probability density function (pdf): 
\begin{equation} 
p_Z(z) = \frac{\gamma_{ggd} z^{\gamma_{ggd}\nu-1}}{\alpha^{\gamma_{ggd}\nu}\Gamma(\nu)} \exp{\left\{-\frac{z}{\alpha}^{\gamma_{ggd}}\right\}}, z\geq0,\alpha,\nu,\gamma_{ggd}>0
\label{eq:GGD}
\end{equation}

To demonstrate robustness to speckle noise, anatomical edges such as bone or layered tissue is models as a square wave $\hat{I}$:

\begin{equation} 
Q(x) = \begin{cases} 
q_l & \text{if } x < a \\
q_u & \text{if } a \leq x < b \\
q_l & \text{if } x \geq b 
\end{cases}
, 0 \leq q_l \ll q_u \leq 255
\end{equation}

where $a,b$ are the transition time steps and $q_l, q_u$ are the lower and upper values. Consequently
\[
(q_l\odot \xi_m \oplus \xi_a) \ll q_u \quad \text{and} \quad 
\begin{aligned}
|q_l - (q_l\odot \xi_m \oplus \xi_a)| &\ll |q_u - q_l|, \\
|q_u - (q_u\odot \xi_m \oplus \xi_a)| &\ll |q_u - q_l|,
\end{aligned}
\]

which therefore indicates that artificial structures could only be created by sampling from the extreme tail of $p_Z$ 
\begin{equation}
P\!\Biggl(\bigcap_{i=1}^{n} \{Z_i > \hat{z}\}\Biggr) = \left(\int_{\hat{z}}^{\infty} p_Z(z)\,dz\right)^n,\quad \hat{z}\gg0,
\end{equation}
which is exponentially unlikely.

Experimental evaluation of the robustness of the algorithm to speckle noise is performed by quantifying the topological feature discrepancies when speckle noise is applied. The speckle noise  applied via Eq.~\ref{eq:speck}, \ref{eq:GGD}, with the $\nu$, $\gamma_{ggd}$ values varied within a range $\{1, 1.5, 2, 2.5, 3\}$, and sampled from the GGD, using Rejection Sampling \cite{neumann1951various}. To evaluate the topological difference, the $q$-Wasserstein distance (with $q=2$), is used to measure the discrepancies between persistence diagrams, derived from the output point cloud \cite{Maria2014TheGL, jmlrpot, Lacombe2018LargeSC}. This measures the topological perturbation under varying noise influence. The q-Wasserstein distance can be defined as $W_q(D,\hat{D}) = \inf_{\varphi:D\rightarrow\hat{D}}\left(\sum_{t\in D}||t-\varphi(t)||^q_\infty\right)^{\frac{1}{q}}$ \cite{Mileyko2011ProbabilityMO}, where $\varphi$ ranges over bijections between the two persistence diagrams $D$ and $\hat{D}$ (affected by speckle noise). $W_q$ is normalised for interpretability.

\subsubsection{Classification evaluation and comparison setup} For the evaluation of the AS classification, the proposed method ($Topol$) is compared to a statistical method that uses the sum of the intensity of a scan line ($Baseline_1$), a statistical method that uses the change in the intensity of the pixels ($Baseline_2$) and a neural network $VGG16$ \cite{Simonyan15}. The $Baseline_1$ method was motivated by \cite{Filho2006SegmentationOC} \cite{Lee2018AutomaticDO}. The use of the VGG16 architecture was motivated by \cite{Melapudi2022ExploitingAP}. $Baselline_2$ is used for the ablation study, the proposed method up to the iterative filtering standard deviation. Confusion matrix analysis is performed, where a positive classification is for AS. The methods are ranked using $F_{\beta} = \frac{(1 + \beta^2) \times PPV \times TPR}{\beta^2 \times PPV + TPR}$, \textcolor{black}{where PPV is a positive predictive value, and TPR is the true positive rate} $\beta=0.5$ to prioritise precision over sensitivity.

The parameter tuning for Topol is done for each fold using the seen subset exclusively. The best performing set of parameters for $Topol$ - found using a grid search - was near identical across the 3 folds $[f=[2,3,4,5,6], \sigma_{u}=1.8, \sigma_{w}=20, \phi_{1}=15, \phi_{2}=20, \epsilon=60]$. With the minor exception of the Gaussian filter where folds 1 and 3 used $\sigma_{v}=0.2$, whilst fold 2 used $\sigma_{v}=0.3$. The scan line threshold, to determine the AS, was constant for all folds $\tau=2$. Although the parameters were chosen using a manually defined grid, this process could easily be performed fully automatically using any hyperparameter optimisation method. 

$Baseline_1$ individually classifies each scan line through scan line summing, with a fixed threshold $\sum_{i=1}^{300}\xi_{sl, i}<\kappa_{Baseline_1}$. The threshold is determined using a receiver operating characteristic curve (ROC). For folds $[1,2,3]$ - $\kappa_{Baseline_1}=[4900,5000,5500]$. \textcolor{black}{The ROC process is repeated for $Baseline_2$}, where for folds $[1,2,3]$ - $\kappa_{Baseline_2}=[325,350,350]$. A deep learning method is implemented to gauge generalisability given data limitations. Linear probing was used with a frozen pretrained VGG16 feature encoder (ImageNet) with a trainable classification head, \textcolor{black}{where the outputs are the classification scores for each scan line in the image.} Training is performed for 10 epochs using Sigmoid cross-entropy. $Topol$, $Baseline_1$ and $Baseline_2$, were tuned using \textcolor{black}{accuracy (ACC).}

Parameters robustness is explored by individually perturbing the calibrated parameters from fold 3 and measuring the effect on the corresponding unseen data. For the parameter study, a maximum perturbation of $20\%$ was deemed suitable for a valid robustness evaluation. The coefficient of variation $=100\times\frac{\sigma_{pert}}{\mu_{pert}}$ is used to measure the impact on performance.

\subsubsection{Robotic Setup}
To extend the AS experiments, a robotic setup is created using the AS output to guide a scanning process - parameters from fold 1 were used. \textcolor{black}{Specifically, AS detection is used to control probe-tissue coupling during the landing, rotation, and the tilting of the probe, optimising probe–tissue (by correcting the in plane rotation through adjustment along the longitudinal axis of the probe and centring the US imaging) contact throughout the scanning process and probe manipulation.} The experiments were carried out using a seven DoF DLR \textsc{Miro}~\cite{Seibold2018}, with a \emph{GE LOGIQ e} US machine and a \emph{GE 12L-RS} linear US probe. The phantom is created using a mixture of GELITA\textregistered GELATINE \emph{Type Ballistic 3} with water and glycerine, with olives and blueberries added for echo diversity \cite{morehouse2007addition}.

Through Cartesian forward kinematics $f_{\mathrm{kin}}:\Qc\rightarrow SE(3) \mid \qq \mapsto \xx=f_{\mathrm{kin}}(\qq)$, the pose of the US probe $\xx\in SE(3)$ can be represented as a function of the generalised configuration coordinates of the robot $\qq$. A Cartesian impedance controller~\cite{imp,Khatib1987} can be constructed to control a robot, by determining desired joint torques $\ttau\in\RRR^{n}$.

\begin{equation}
    \ttau = \JJ\tran(\qq)[\KK(\xx_d-\xx)+\DD(\qq)(\dot{\xx}_d-\dot{\xx})],
    \label{eq:imp_law}
\end{equation}

where $\xx$ and $\dot{\xx}$ are the current pose and velocity. This forms a spring-damper system with stiffness and damping parameters $\KK,\DD(\qq)\in\RRR^{6\times6}$. Enabling control of pose $\xx_d$ and velocity $\dot{\xx}_d$ of a US probe. The geometrical Jacobian $\JJ(\qq)\in\RRR^{6\times 7}$ is given by $\dot{\xx}=\begin{bmatrix}\vv & \oomega\end{bmatrix}\tran=\JJ(\qq)\dot{\qq}$.

\begin{align}
    \label{eq:m1}
    m_1 &= \frac{1}{width} \sum_{sl=1}^{width} (1 - \xi_{sl}), \quad m_1\in[0; 1]\\
    \label{eq:m2}
    m_2 &= \frac{1}{width/2} \left(\sum_{sl=width/2}^{width} \xi_{sl} - \sum_{sl=1}^{width/2} \xi_{sl}\right), \quad m_2\in[-1; 1],
\end{align}
Two metrics are defined to measure the coupling - $m_1$ is the coverage of contact, $m_2$ is the symmetry of contact. These metrics can guide the scan by defining desired values $m_{1,d}$ and $m_{2,d}$ and mapping the error to a desired Cartesian velocity \textcolor{black}{expressed in the probe's coordinate frame, enabling longitudinal correction of the contact}.

\begin{equation}\label{eq:xd}
\dot{\xx}_d=
    \begin{bmatrix}
        \vv_{d,m} \\ \oomega_{d,m} + \oomega_{d,t}+\oomega_{d,r}
    \end{bmatrix}=
    \begin{bmatrix}
        (m_{1,d}-m_1) 
        \begin{bmatrix}
            0 & 0 & k_1
        \end{bmatrix}\tran \\ 
        (m_{2,d}-m_2)
        \begin{bmatrix}
            k_2 & 0 & 0
        \end{bmatrix}\tran 
        + \oomega_{d,t}+\oomega_{d,r}
    \end{bmatrix},
\end{equation}

with two gains $k_1=0.005~\nicefrac{\text{m}}{\text{s}}$ and $k_2=0.05~\nicefrac{\text{rad}}{\text{s}}$. The translation velocity is defined by $\vv_{d,m}$, while the rotation velocities are defined by $[\oomega_{d,m}, \oomega_{d,t}, \oomega_{d,r}]$. \textcolor{black}{Here the probe coordinate frame is defined so that the y axis aligns with the scanning direction, the z axis with the longitudinal axis of the probe and the x axis orthogonal to both.} The integration of Eq.~\eqref{eq:xd} gives the desired pose $\xx_d$. If no trajectory is defined ($\oomega_{d,t}=\oomega_{d,r}=[0,0,0]$), so, for example, when the probe is landing or maintaining perpendicular contact. Using $\dot{\xx}_d$, the robot will automatically adjust the position of the probe along the longitudinal axis until $m_1=m_{1,d}$, and adapt the orientation until ($m_2=m_{2,d}=0$). To perform rotation and tilt, $\oomega_{d,t}=\begin{bmatrix}0&*&0\end{bmatrix}\tran$ and $\oomega_{d,r}=\begin{bmatrix}0&0&*\end{bmatrix}\tran$ accordingly, where $*$ is defined by the operator.

\subsection{Medical Phantom Data}

\subsubsection{Data curation} Controlled evaluation of confidence is difficult, as optimal scans may produce subpar images due to unavoidable factors such as forced positioning, \textcolor{black}{e.g., limited space within a cavity or the location of the pathological tissue.} Data was collected for relative evaluation, which involved tilting the probe from -90\textdegree \ $>$ 0\textdegree (\textcolor{black}{aligned with the normal of the tissue surface}) $>$ +90\textdegree \ with respect to the normal vector of the tissue surface. Data was captured using a Zonare Z.one US system with a C4-1 transducer. 
Four scans were taken using the FAST / acute abdomen Phantom. Scans 1 and 2 from the abdomen, scan 3 from the rib cage, and scan 4 from the kidney. A fifth scan was taken using a Kezlex brain phantom. Frames were extracted at $10Hz$. For generalisation evaluation of AS classification, 449 frames are extracted and annotated.

\begin{figure}[!htbp]
\centering
\includegraphics[width=\textwidth]{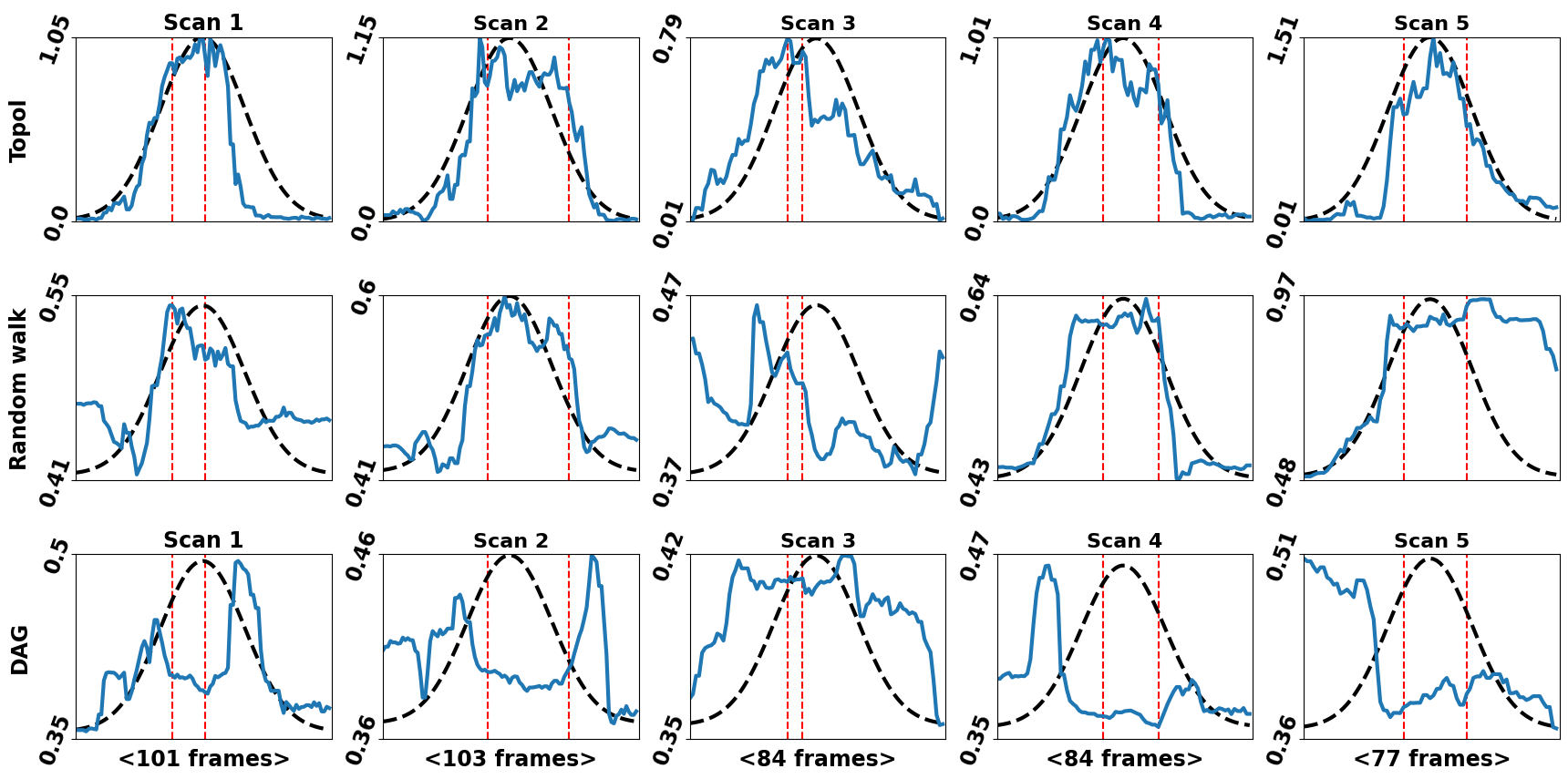}
\caption{The per frame average confidence vector outputs - x axis = frames and y axis = average confidence. The blue lines show the method's outputs, $Topol$ top, $Random walk$ middle and $DAG$ bottom row. The black dotted lines are the scaled normal curve. The dotted red lines are where the annotators perceived the best images. What can be observed is that the output of $Topol$ follows the desired trajectory, closer than the other methods.}
\label{fig:pose}
\end{figure}

\begin{figure}[!htbp]
    \centering
    \includegraphics[width=\linewidth]{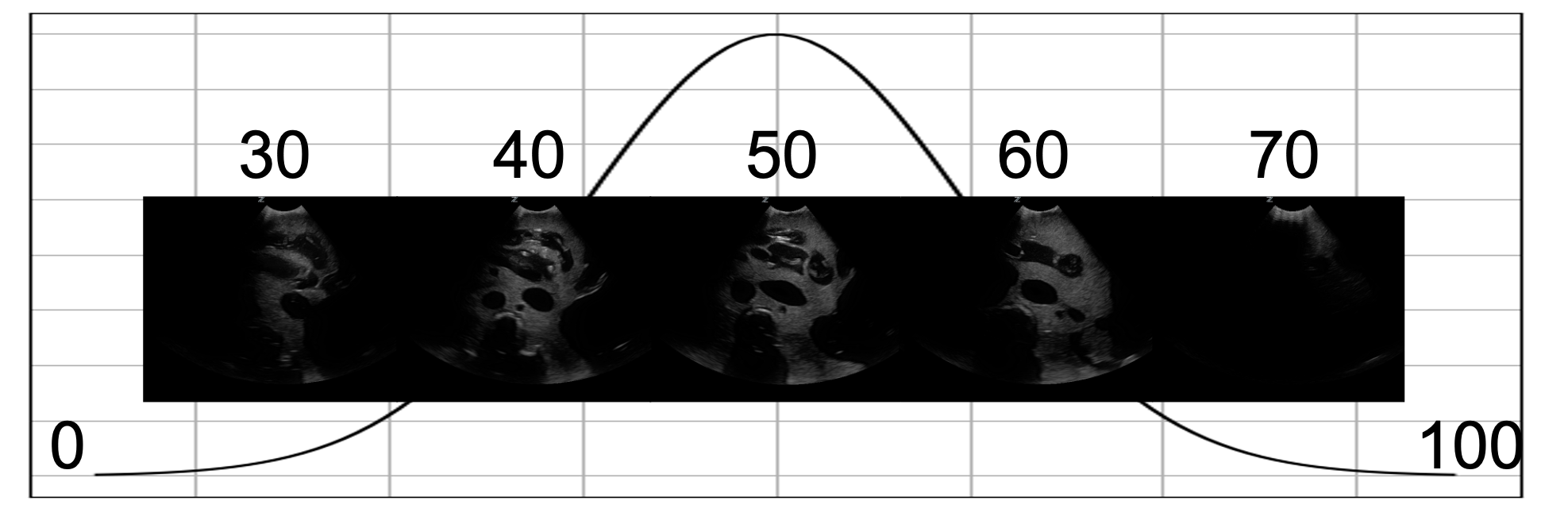}
    \caption{The image feedback at different frames from Scan 1 - abdominal. The optimal image is shown at frame 50.}
    \label{fig:posediff}
\end{figure}

\subsubsection{Confidence map evaluation and comparison setup} The proposed method is compared with the $Random walk$ \cite{Karamalis2012USCM} and $DAG$ \cite{Hung2020UltrasoundCM} methods, using their official codes and default parameters. 
The $Topol$ parameters tuned on fold 1 of the AS classification were used for this experiment. Due to the way the data was collected, the target confidence vector is defined as a normal distribution across the frames.  Based on \cite{Virga2016AutomaticFR}, $mean(\hat{M})$ is used to determine the confidence in the perceptual salience of the visible anatomy per frame. The evaluation is performed using the normalised root mean squared error, $(NRMSE) =\frac{\sqrt{\frac{1}{n} \sum_{i=1}^{n} (T_{i} - \hat{T}_{i})^2}}{\max(T) - \min(T)}, \hat{T}_{i}=mean(\hat{M}_{i})$, where $T$ is the target confidence vector, the black dotted lines in Fig.~\ref{fig:pose}, fitted to the max/min of $\hat{T}$; $n$ is the frame.  Fig.~\ref{fig:posediff} is from Scan 1. The best image is the middle frame, and the black curve is representative of the assumed confidence trajectory. To confirm this, annotations were made to highlight the optimal frame range, highlighted with the red dotted lines in Fig.~\ref{fig:pose}. 

\subsubsection{Classification generalisability}
To test the generalisability and performance of each algorithm on out-of-distribution data, AS scanline classification experiments were extended to phantom data (different machine, anatomy and image texture). Each method was evaluated three times, using the three sets of parameters calibrated for the three folds described in Sec.~\ref{sec_slc}, and the average results across the three folds were recorded.

\section{Results and Discussion}\label{sec_rd}

\begin{figure}[!htbp]
\centering
\includegraphics[height=2cm, width=0.33\columnwidth]{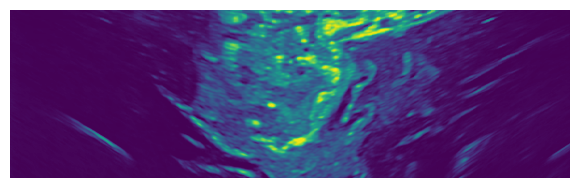}
\includegraphics[height=2cm, width=0.33\columnwidth]{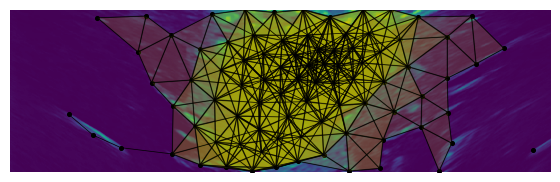}
\includegraphics[height=2cm, width=0.33\columnwidth]{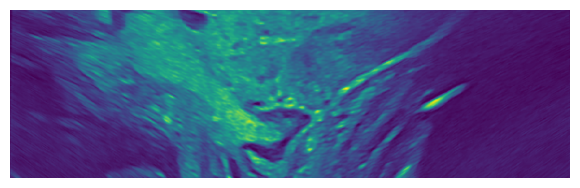}
\includegraphics[height=2cm, width=0.33\columnwidth]{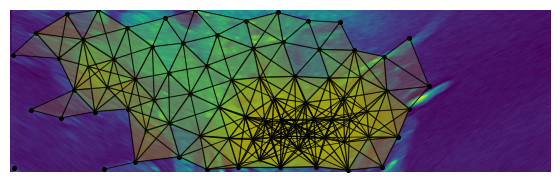}
\includegraphics[height=2cm, width=0.33\columnwidth]{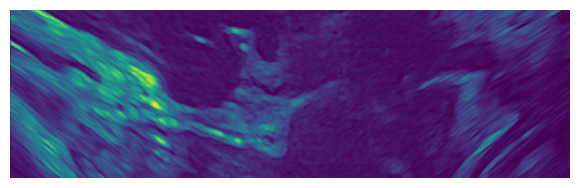}
\includegraphics[height=2cm, width=0.33\columnwidth]{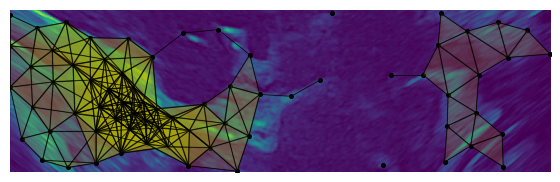}
\includegraphics[height=2cm, width=0.33\columnwidth]{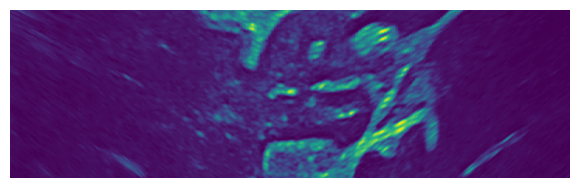}
\includegraphics[height=2cm, width=0.33\columnwidth]{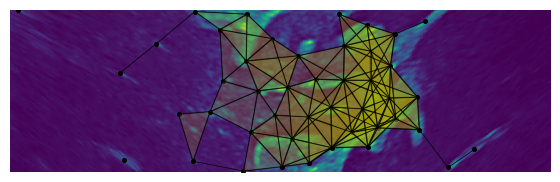}
\caption{ Sample simplicial complex outputs. These complexes indicate areas of significant spatial intensity variation within the ultrasound images. Areas where there are dense complexes, correlate with high confidence in tissue visibility; correlating with a confidence map output of higher perceptual quality.}
\label{fig:simplex_plots}
\end{figure}

The evaluation was carried out primarily using a Python implementation; results are shown in Fig.~\ref{fig:simplex_plots}. The method was also implemented with minimal optimisation using C++, which ran at 20Hz, indicating real-time executability.

\subsection{Noise Robustness}

The q-Wasserstein distances, after applying the set speckle noise $\{1, 1.5, 2, 2.5, 3\} \times \{1, 1.5, 2, 2.5, 3\}$ for $\nu\times\gamma_{ggd}$, measured an average result of 0.097. This result gives evidence of the robustness of the topological construction under varying speckle noise. The topological differences are largely uniform across all distributions.

\subsection{Acoustic Shadow Classification}\label{sec_slc}

\begin{table}[!htbp]
\small
\centering
\caption{Scan line classification results from the intraoperative data. A positive sample corresponds to a scan line affected by AS. Highlighted in grey is the best score for each metric. The results have been averaged over the 3 folds.}
\begin{tabular}{|l| c| r|r|r|r|} 
\hline
3F Avg  & Algo & $ACC$ & $TNR$ & $TPR$ & $PPV$
\\
\hline 
& $Baseline_1$ & 0.87 & 0.88 & 0.86 & 0.86 \\
& $Baseline_2$ & 0.88 & 0.91 & 0.85 & 0.90 \\
\multirow{-2}{*}{Seen/Train} & $VGG16$ & \cellcolor{blue!25}0.90 & 0.92 & \cellcolor{blue!25}0.90 & 0.90  \\
& $Topol$ & 0.89 & \cellcolor{blue!25}0.95 & 0.83 & \cellcolor{blue!25}0.93 \\

\hline & $Baseline_1$ & 0.84 & 0.81 & \cellcolor{blue!25}0.89 & 0.79\\
& $Baseline_2$ & \cellcolor{blue!25}0.87 & 0.85 & 0.87 & 0.83 \\
\multirow{-2}{*}{Unseen/Test} & $VGG16$ & 0.83 & 0.82 & 0.84 & 0.80 \\
& $Topol$ & \cellcolor{blue!25}0.87 & \cellcolor{blue!25}0.90 & 0.84  & \cellcolor{blue!25}0.87 \\
\hline 
\end{tabular}
\label{tab:as_results}
\end{table}
\begin{table}[t]
\small
\centering

\caption{The rankings (R\#) of the methods using the $F_{\beta}$ scores. The results are a product of averaging across the 3 folds.}
\centering
\begin{tabular}{|l|r|r|r|r|} 
\hline
3F Avg & \cellcolor{blue!25}R1 & R2 & R3 & R4 
\\
\hline 

 & \cellcolor{blue!25}$Topol$ & $VGG16$ & $Baseline_2$ & $Baseline_1$ \\
\multirow{-2}{*}{Seen/Train}&   \cellcolor{blue!25}0.908 & 0.900 &  0.890 & 0.860 \\

\hline  & \cellcolor{blue!25}$Topol$ & $Baseline_2$ & $VGG16$ & $Baseline_1$ \\
\multirow{-2}{*}{Unseen/Test}& \cellcolor{blue!25}0.864 & 0.838 & 0.808 & 0.808 \\
\hline 
\end{tabular}
\label{tab:as_results_fbeta}
\end{table}

\begin{figure}[!htbp]
\centering 
\begin{subfigure}{0.3\columnwidth}
  \includegraphics[width=\linewidth]{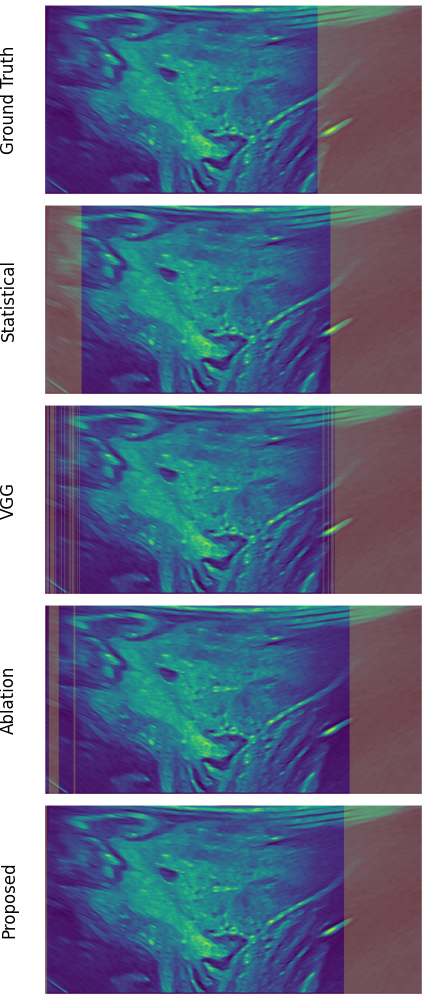}
  \label{fig:4}
\end{subfigure}
\begin{subfigure}{0.3\columnwidth}
  \includegraphics[width=\linewidth]{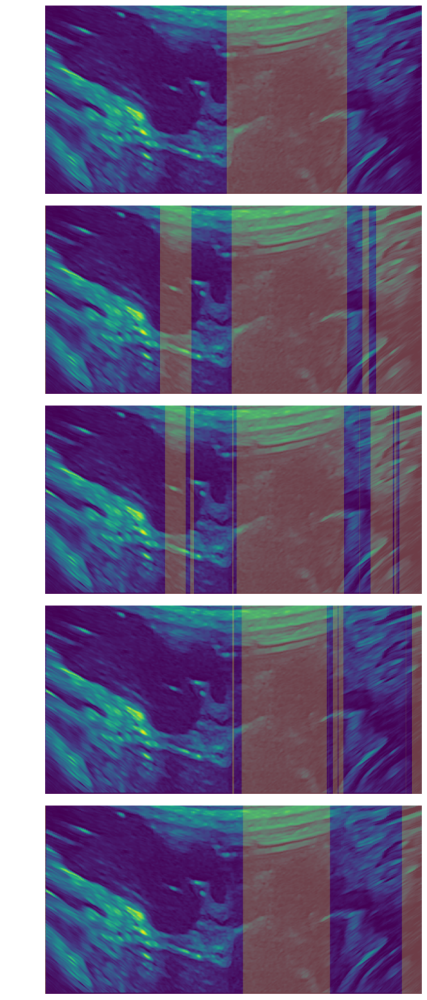}
  \label{fig:4}
\end{subfigure}
\vspace{-1.5em}
\caption{AS detection on two US images. The top row shows the annotation, the second is from $Baseline_1$, the third is from $VGG16$ and the bottom is from Topol. The orange areas are detected AS.}
\label{fig:scanline_as}
\end{figure}

The results are presented in Tab.~\ref{tab:as_results},~\ref{tab:as_results_fbeta} \textcolor{black}{where TNR is the true negative rate}. Outputs are shown in Fig.~\ref{fig:scanline_as}. Firstly, it can be determined that $Topol$ is the most robust method for distribution shift with more consistent performance between seen and unseen data. The results produced by $Baseline_1$ show that overreliance on the raw pixel intensities leads to oversensitivity. By producing results superior to $Baseline_1$, $Baseline_2$ justifies the iterative filtering process, demonstrating that identifying areas of dynamic change is a strong indicator of where there is AS. $VGG16$ shows achieves inadequate generalisation. From the $F_{\beta}$ ranks, it is conclusive that $Topol$ produces the ideal behaviour. 

\subsubsection{Generalisability Study}

\begin{table}[!htbp]
\small
\centering
\caption{The rankings (R\#) of the methods' performances on the generalisability test on the probe pose analysis data, using the $F_{\beta}$ scores. For all scans, $Topol$ achieves the best rank.}
\setlength{\extrarowheight}{1.5pt}

\begin{tabular}{|l|r|r|r|r|} 
\hline
 & \cellcolor{blue!25}R1 & R2 & R3 & R4 
\\
\hline

 & \cellcolor{blue!25}$Topol$ & $Baseline_2$ & $Baseline_1$ & $VGG16$ \\
\multirow{-2}{*}{Scan 1} &   \cellcolor{blue!25}0.923 & 0.842 &  0.807 & 0.694 \\
\hline
 & \cellcolor{blue!25}$Topol$ & $Baseline_2$ & $Baseline_1$ & $VGG16$ \\
\multirow{-2}{*}{Scan 2}&   \cellcolor{blue!25}0.920 & 0.883 &  0.789 & 0.740 \\
\hline
 & \cellcolor{blue!25}$Topol$ & $Baseline_2$ & $Baseline_1$ & $VGG16$ \\
\multirow{-2}{*}{Scan 3}&   \cellcolor{blue!25}0.833 & 0.789 &  0.745 & 0.377 \\
\hline
 & \cellcolor{blue!25}$Topol$ & $VGG16$ & $Baseline_2$ & $Baseline_1$ \\
\multirow{-2}{*}{Scan 4}&   \cellcolor{blue!25}0.891 & 0.847 &  0.797 & 0.753 \\
\hline
 & \cellcolor{blue!25}$Topol$ & $VGG16$ & $Baseline_2$ & $Baseline_1$ \\
\multirow{-2}{*}{Scan 5}&   \cellcolor{blue!25}0.830 & 0.777 &  0.686 & 0.643 \\
\hline 
\end{tabular}
\label{tab:as_results_fbeta_scan}
\end{table}

The results of \textcolor{black}{the phantom dataset experiments} are shown in Tab.~\ref{tab:as_results_fbeta_scan}. For all scans, $Topol$ achieves the best performance. The ranking of $Baseline_2$ again highlights the advantage of using the intensity change rather than the raw intensity. The inconsistent performance of $VGG16$ highlights the unreliability of deep learning methods when only small datasets are available.

\subsubsection{Parameter Perturbation Study}\label{sec:param_pert}

\begin{table}[!htbp]
    \centering
    \caption{Parameter perturbation, performance drift measured via coefficient of variation values (\%).}
    \begin{tabular}{lcccc}
        \toprule
        Parameter & ACC & TNR & TPR & PPV \\
        \midrule
        $f$       & 0.431  & 1.928  & 2.204  & 1.749  \\
        $\sigma_u$ & 0.320  & 1.191  & 1.455  & 1.011  \\
        $\sigma_v$ & 0.455  & 2.380  & 2.290  & 2.151  \\
        $\gamma$  & 0.982  & 4.499  & 3.986  & 3.925  \\
        $\phi_1$  & 0.622  & 1.683  & 1.453  & 1.579  \\
        $\phi_2$  & 0.512  & 1.150  & 1.550  & 1.031  \\
        $\sigma_w$ & 0.358  & 0.277  & 0.635  & 0.347  \\
        $\epsilon$ & 2.270  & 8.663  & 5.986  & 6.306  \\
        \bottomrule
    \end{tabular}
    \label{tab:perty}
\end{table}

\setlength{\fboxrule}{1pt}  
\setlength{\fboxsep}{0pt}

\begin{figure}[!htbp]
\centering
\includegraphics[width=0.32\linewidth]{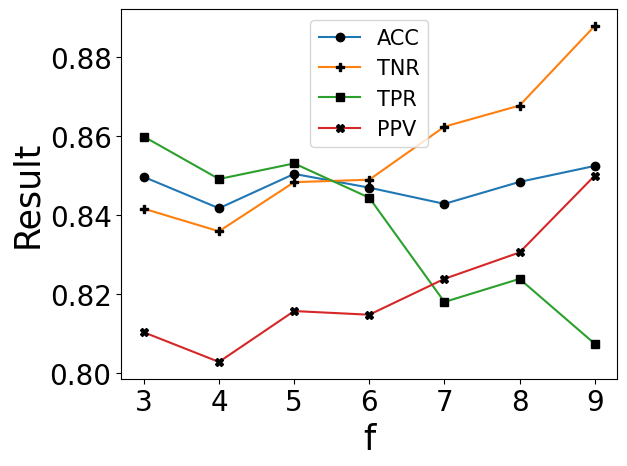}
\includegraphics[width=0.32\linewidth]{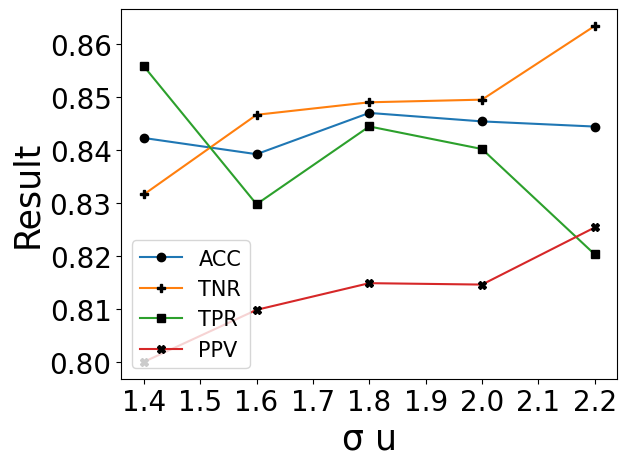}
\includegraphics[width=0.32\linewidth]{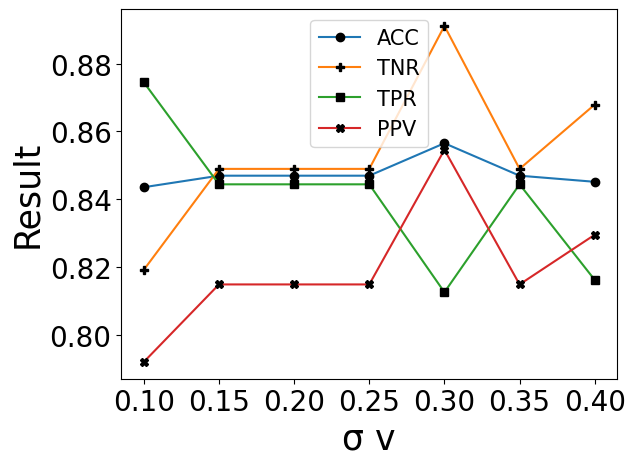}
\includegraphics[width=0.32\linewidth]{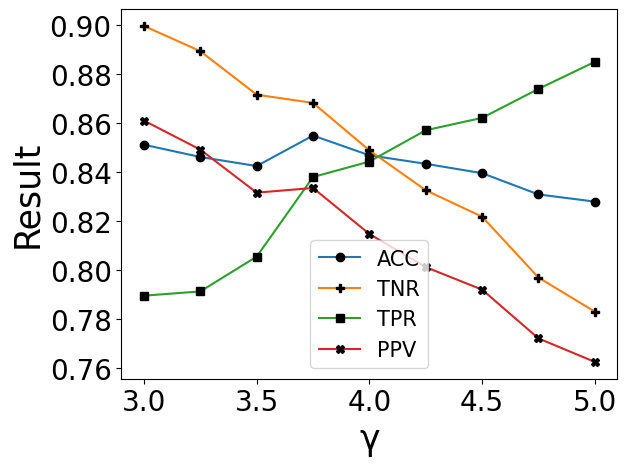}
\includegraphics[width=0.32\linewidth]{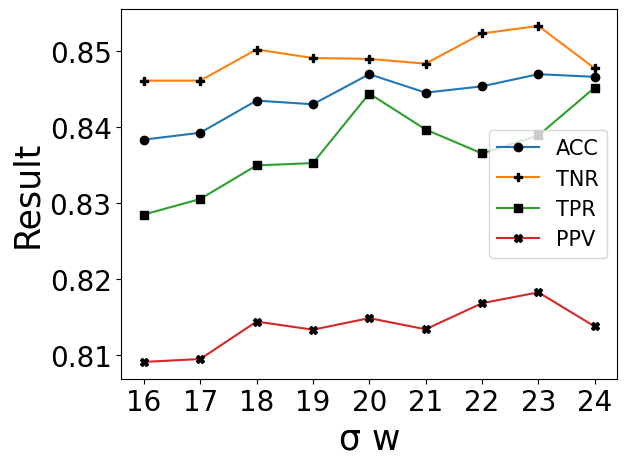}
\includegraphics[width=0.32\linewidth]{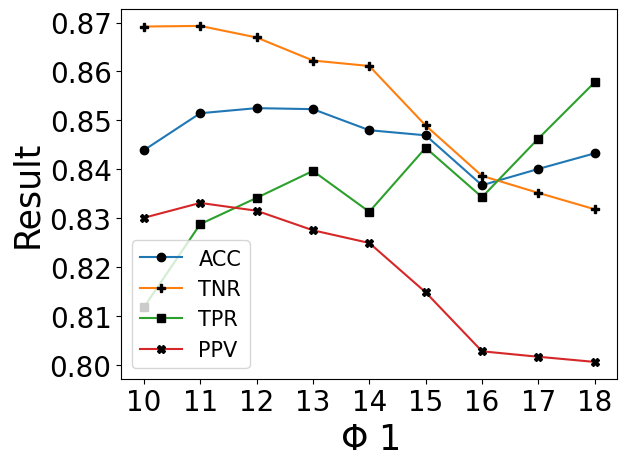}
\includegraphics[width=0.32\linewidth]{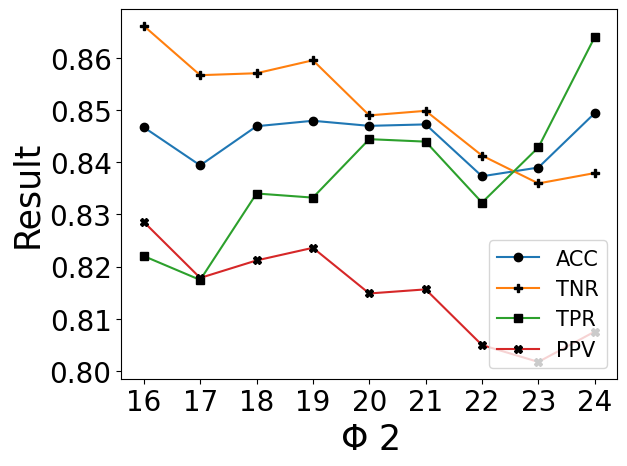}
\includegraphics[width=0.32\linewidth]{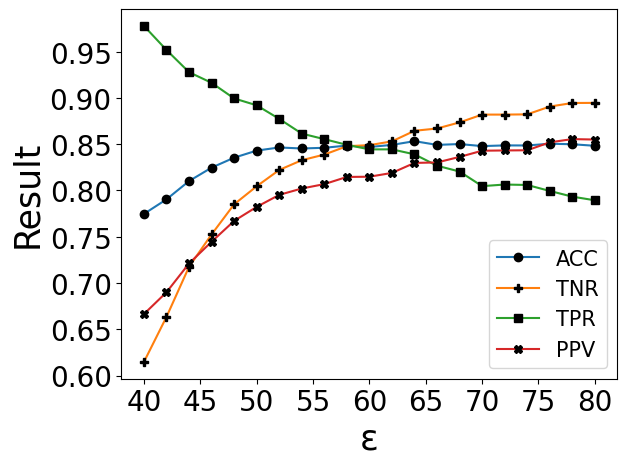}
\caption{Graphs displaying the effect of individually varying the parameters on the performance of the whole system - using the unseen fold 3 data. blue lines = ACC, orange lines = TNR, green lines = TPR and red lines = PPV.}
\label{fig:pert}
\end{figure}

The results of this experiment are shown in Fig.~\ref{fig:pert}. It is observed that the overall performance is only marginally affected by individually drifting the parameters, with most metrics remaining above 0.8. The coefficient of variation values (\%) of the performance drift are shown in Tab.\ref{tab:perty}. This is evidence of the robustness of the design of the algorithm, as the performances do not vary largely under parameter uncertainty. We note that the $\gamma$ value that was decided was initially based on intuition from the iterative filtering process. In particular, calculating the overall standard deviation $\sqrt{5}\times1.8\approx4.02$ yields a value close to $\gamma$=4. This result suggests a potential correlation between the iteration and the optimal threshold choice.

\subsection{Robotics results}

\begin{figure}[!htbp]
\centering
\includegraphics[width=0.8\linewidth]{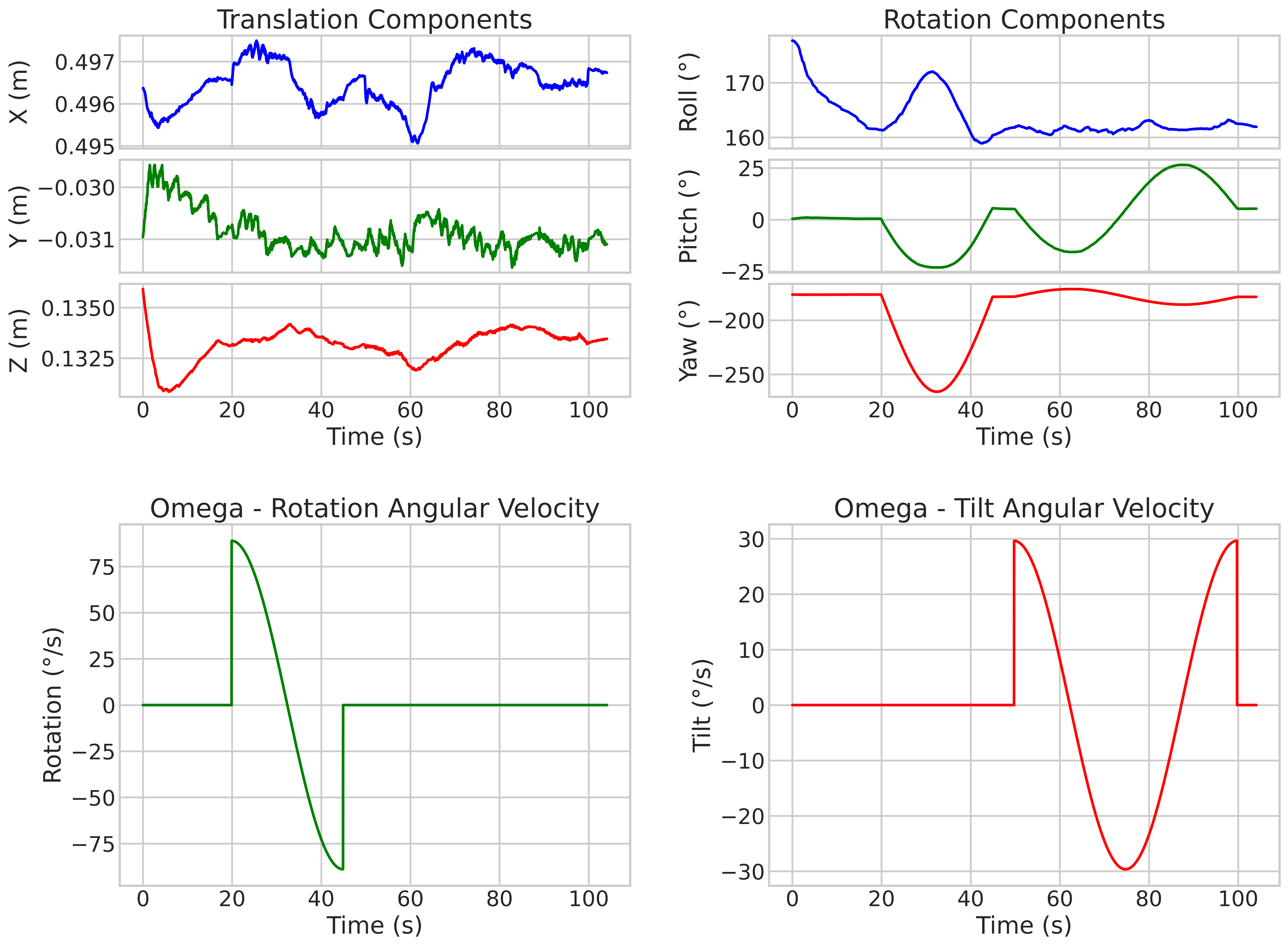}
\caption{\textcolor{black}{The top row shows the real robot movement, the bottom row shows the desired angular velocities (trajectories).}}
\label{fig:robbase}
\end{figure}

\begin{figure}[!htbp]
\centering
\includegraphics[width=0.8\linewidth]{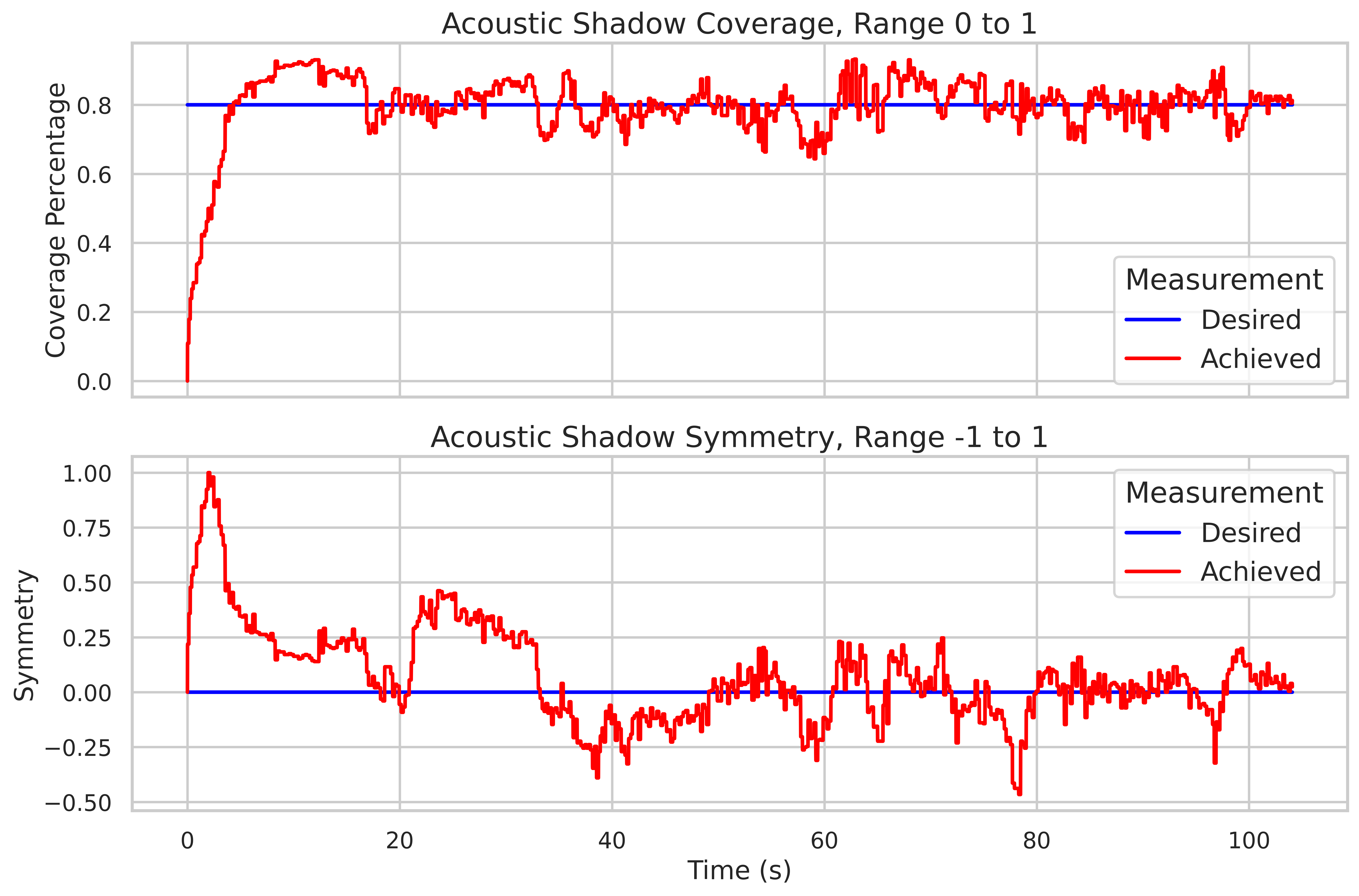}
\caption{Acoustic shadow behaviour and optimisation for robot scan.}
\label{fig:robacsh}
\end{figure}

The sequence of robotic action is first the landing and stabilisation of the probe on the phantom surface, followed by a $\pm90$\textdegree \ rotation, followed by a $\pm30$\textdegree \ tilt. The \textcolor{black}{actual position and orientation, measured via the homogeneous transformation matrix in the robot base frame}, as well as the \textcolor{black}{corresponding position and orientation} commands (desired trajectories) of the probe, are shown in Fig.~\ref{fig:robbase}. Of the translation components, Z is of greatest interest, as the scan was held on a single spot, and Z was the main translation for correction. The real homogeneous transformations matrices show that the robot correctly executed the motion without large deviation. The AS results for $[m_1, m_2]$ are shown in Fig.~\ref{fig:robacsh}. It can be determined that AS analysis enabled the control framework to successfully anchor the contact to the desired coverage and symmetry.

\subsection{Confidence Map Behaviour}\label{conf_exp}

Fig.~\ref{fig:pose} shows that for all scans, the behaviour of $Topol$ is closer to what is expected. The range of $\hat{T}$ values produced by $Topol$ can also be considered more intuitive than $Random walk$ and $DAG$ as the confidence starts near 0 and peaks around 1. Quantitative analysis is also performed and shown in Tab.~\ref{tab:pose_tab}, produced by measuring the NRMSE of $\hat{T}$ (blue line) versus $T$ (black dotted line). For each scan, the proposed method produces a significantly better NRMSE score than the other methods. 

\begin{table}[!htbp]
\small
\centering
\setlength\tabcolsep{3pt}
\caption{The method's NRMSE results are shown for each scan. In all cases, $Topol$ produces the best score. }

\begin{tabular}{|l|r|r|r|r|r|} 
\hline
Method & Scan 1 & Scan 2 & Scan 3 & Scan 4 & Scan 5 
\\ 
\hline 
$Topol$ & \cellcolor{blue!25}0.18 & \cellcolor{blue!25}0.13 & \cellcolor{blue!25}0.28 & \cellcolor{blue!25}0.14 & \cellcolor{blue!25}0.11\\
$Random walk$ & 1.85 & 0.94 & 4.79 & 0.73 & 0.87 \\
$DAG$ & 2.40 & 4.96 & 6.08 & 5.12 & 4.15 \\
\hline 
\end{tabular}
\label{tab:pose_tab}
\end{table}

\section{Conclusion}\label{sec_conc}

This paper introduces a novel method for evaluating the quality of US feedback. An evaluation of AS scan line classification and confidence behaviour has been performed. In both cases, the proposed method outperforms the compared methods. Further experiments show that this method is capable of guiding a robotic arm in performing an autonomous scan. The explicit evaluation of US probe-tissue coupling is vital for ensuring safe scanning. This method provides a novel perspective on how this can be approached, creating an avenue of future work which can be beneficial for both robotic US and supporting trainee operators.

\backmatter

\bmhead{Acknowledgements}

This project was supported by the UK Research and Innovation (UKRI) Centre for Doctoral Training in AI for Healthcare (EP/S023283/1), the Royal Society (URF$\setminus$R$\setminus$201014), the NIHR Imperial Biomedical Research Centre.

\section*{Declarations}

\begin{itemize}
\item Conflict of interest: The authors declare no conflict of interest.
\end{itemize}

\bibliography{inbib}

\end{document}